\documentclass[12pt]{article}
\newcommand{\eps}{\varepsilon}
\newcommand{\Li}{\mathrm{Li}_2}
\newcommand{\dd}{{\mathrm d}}
\newcommand{\vecc}[1]{\mbox{\boldmath $#1$}}

\title{LABSMC: Monte Carlo event generator for large-angle 
Bhabha scattering}

\author{A.B. Arbuzov\thanks{on leave of absence from 
Joint Institute for Nuclear Research, Dubna, Russia.}
}

\date{}

\begin{document}
\maketitle

\begin{itemize}
\item[$ $]
        {\em Dipartimento di Fisica, Universit\`a di Torino,
             and INFN group, \\
             via Guiria 1, I-10125 Torino, Italy \/} \\
        {\tt e-mail: arbuzov@to.infn.it}
\end{itemize}

\begin{abstract}
A Monte Carlo event generator is presented. An original algorithm
is developed to simulate electron--positron scattering at energies 
and momentum transferred much more than the electron mass. 
The first-order electroweak radiative corrections are included 
completely.  Higher order corrections are taken into account by 
means of electron structure functions. 
\\[.2cm] \noindent
\end{abstract}

\section{Introduction}

The process of electron--positron (Bhabha) scattering was studied both 
theoretically and experimentally since many years~\cite{Bhabha}. 
It has almost pure 
electrodynamical nature and therefore could be described with very high 
precision by means of perturbative QED. The process is commonly used
at $e^+e^-$ colliders for luminosity measurements, because it has a
large cross section and can be measured very accurately. The modern
experimental technique of luminosity measurements reaches the one per mille
level of accuracy, or even better, as at LEP1~\cite{LEP1}. 
This is a challenge for the theory. At the Born and one--loop levels the 
process was investigated
in detail (see papers \cite{l3,l4,l5} and references therein), 
taking into account both QED and electroweak effects.
The radiative corrections in
the first order in the fine structure constant $\alpha$ become
insufficient now, one has to take into account higher order effects.
In order to meet the requirements of experiments one has also to
implement the results of analytical calculations into a Monte Carlo
event generator. 

In this note we present an event generatorr, based on 
the approach~\cite{fed}, how to merge the complete
${\cal O}(\alpha)$ result with the leading logarithmic corrections
in higher orders. 
The structure of the paper is as follows.
In the next section we give the master formula for description
of large--angle Bhabha scattering and decompose it into 13 parts
of different kinematics. The consequent structure of the Monte Carlo code
is described in Sec.~3. The main options and parameters of the program
are presented in Sec.~4. Numerical results and the precision achieved
are discussed in the Conclusions.

\section{The master formula}

The reaction 
\begin{eqnarray}
e^-(p_1)\ +\ e^+(p_2)\ \to\ e^-(p_1')\ +\ e^+(p_2')\ +\ (n\gamma)
\end{eqnarray}
will be considered in the centre--of--mass reference frame of the 
incoming particles.

Let us start with the master formula in the form as given in
paper~\cite{JHEP97a}:
\begin{eqnarray} \label{master}
&& \frac{\dd\sigma^{e^+e^-\to e^+e^-(\gamma)}}{\dd\Omega_-}
= \int\limits_{\bar{z}_1}^{1}\dd z_1\;\int\limits_{\bar{z}_2}^{1}\dd z_2\;
{\cal D}(z_1){\cal D}(z_2)
\frac{\dd\tilde{\sigma}_{0}(z_1,z_2)}{\dd\Omega_-}
\left(1+\frac{\alpha}{\pi}K_{SV}\right)\Theta \nonumber 
\\ \nonumber && \qquad
\times \int\limits^{Y_1}_{y_{\mathrm{th}}}\frac{\dd y_1}{Y_1}\;
\int\limits^{Y_2}_{y_{\mathrm{th}}}\frac{\dd y_2}{Y_2}\;
{\cal D}(\frac{y_1}{Y_1}){\cal D}(\frac{y_2}{Y_2})
\\ \nonumber
&& \qquad + \frac{\alpha}{\pi}\int\limits_{\Delta}^{1}\frac{\dd x}{x}
\Biggl\{\biggl[\left(1-x+\frac{x^2}{2}\right)\ln\frac{\theta_0^2(1-x)^2}{4}
+ \frac{x^2}{2}\biggr]\, 2 \,
\frac{\dd\sigma_0^{\mathrm{Born}}}{\dd \Omega_-}
\\ \nonumber && \qquad
+ \biggl[\left(1-x+\frac{x^2}{2}\right)\ln\frac{\theta_0^2}{4}
+ \frac{x^2}{2}\biggr]
\Biggl[\frac{4\alpha^2}{s(1-x)^2[2-x(1-c)]^4} \\ \nonumber && \qquad \times
\left(\frac{3-3x+x^2+2x(2-x)c+c^2(1-x+x^2)}{1-c}\right)^2
\\ \nonumber && \qquad
+ \frac{4\alpha^2}{s[2-x(1+c)]^4}
\left(\frac{3-3x+x^2-2x(2-x)c+c^2(1-x+x^2)}{1-c}\right)^2 \Biggr]
\Biggr\}\Theta
\\ && \qquad \label{bhabha}
- \frac{\alpha^2}{4s}
\left(\frac{3+c^2}{1-c}\right)^2
\frac{8\alpha}{\pi}\ln(\cot\frac{\theta}{2})
\ln\frac{\Delta\eps}{\eps}
\; + \; \frac{\alpha^3}{2\pi^2s}\!\!\!\!\!
\int\limits_{\stackrel{k^0>\Delta\eps}{\pi-\theta_0>\theta>\theta_0}}
\!\!\!\!\!\! \frac{WT}{4}\;\Theta \frac{\dd \Gamma}{\dd \Omega_-},
\\ \nonumber &&
Y_1 = \frac{2z_1z_2}{z_1+z_2-c(z_1-z_2)}\, , \qquad
Y_2 = \frac{z_1^2+z_2^2-(z_1^2-z_2^2)c}{z_1+z_2-c(z_1-z_2)}\, , 
\\ \nonumber &&
\bar{z}_1 = \frac{y_{\mathrm{th}}(1+c)}{2-y_{\mathrm{th}}(1-c)}\, ,
\qquad\qquad\quad
\bar{z}_2 = \frac{z_1y_{\mathrm{th}}(1-c)}{2z_1-y_{\mathrm{th}}(1+c)}\, .
\end{eqnarray}
Step functions $\Theta$ represent any possible cuts on the phase
space of the corresponding variables.
By $K_{SV}$ we denoted the so-called $K$-factor\footnote{The last
term in the square brackets of the expression for the $K$-factor
in Ref.~\cite{JHEP97a} is incorrect.}, 
which comes
from virtual and soft radiative corrections, and therefore it
can be factorized at the Born cross section $\dd\tilde{\sigma}_{0}$,
\begin{eqnarray}
K_{SV} &=& -1-2\Li(\sin^2\frac{\theta}{2}) + 2\Li(\cos^2\frac{\theta}{2})
+ \frac{1}{(3+c^2)^2}\biggl[\frac{\pi^2}{3}(2c^4 - 3c^3 - 15c)
\nonumber \\ \nonumber
&+& 2(2c^4 - 3c^3 + 9c^2 + 3c + 21)\ln^2(\sin\frac{\theta}{2})
- 4(c^4+c^2-2c)\ln^2(\cos\frac{\theta}{2}) \\ \nonumber
&-& 4(c^3+4c^2+5c+6)\ln^2(\tan\frac{\theta}{2})
+ 2(c^3-3c^2+7c-5)\ln(\cos\frac{\theta}{2}) \\
&+& 2(3c^3+9c^2+5c+31)\ln(\sin\frac{\theta}{2}) \biggr].
\end{eqnarray}
The {\it shifted} ({\it boosted}) Born cross
section (with vacuum polarization effect taken into account) reads
\begin{eqnarray}
\frac{\dd\widetilde{\sigma}(z_1,z_2)}{\dd\varphi\dd c} 
&=& \frac{4\alpha^2}{sa^2}\biggl[ \frac{a^2+z_2^2(1+c)^2}
{(1-\Pi(\tilde{t}))^22z_1^2(1-c)^2} + \frac{z_1^2(1-c)^2+z_2^2(1+c)^2}
{|1-\Pi(\tilde{s})|^22a^2} \nonumber \\
&-& {\mathrm{Re}}\;\frac{z_2^2(1+c)^2}
{(1-\Pi(\tilde{t}))(1-\Pi(\tilde{s}))^* az_1(1-c)} \biggr], \\ \nonumber
\tilde{t} &=& - \frac{1}{2}sz_1Y_1(1-c), \qquad 
\tilde{s} = sz_1z_2, \qquad Y_1 = \frac{2z_1z_2}{a}\, , \\ \nonumber
a &=& z_1 + z_2 - (z_1-z_2)c.
\end{eqnarray}
For detailed notation and the derivation look in Ref.~\cite{JHEP97a}.
The current version of the code includes also the contributions of 
$Z$-exchange and $Z$-$\gamma$ interference as well as the relevant
set of the first order weak radiative corrections to be described
elsewhere.

We quoted the complete expression; now we are going
to discuss it and decompose into a form, suitable for an
event generator.

The first term of Eq.~(\ref{master}) is written in the form of
the Drell--Yan cross section. The leading logarithmic corrections 
are accounted by means of the ${\cal D}$-functions, which are the
kernel functions of the Dokshitzer--Gribov--Altarelli--Parisi--Lipatov
evolution equations.
The rest supplements the sub--leading terms, which come from the 
straightforward calculations in the ${\cal O}(\alpha)$ order. 

The non--singlet electron structure function is expanded in the series in
$\alpha$:
\begin{eqnarray}
{\cal D}(z) &=& \delta(1-z) + \frac{\alpha}{2\pi}(L-1)P^{(1)}(z)
+  \biggl(\frac{\alpha}{2\pi}\biggr)^2\frac{(L-1)^2}{2!}P^{(2)}(z) 
+ \dots \ , \\ \nonumber
P^{(1,2)}(z) &=& \lim_{\Delta\to 0}\biggl\{\delta(1-z)P^{(1,2)}_{\Delta}
+ \Theta(1-\Delta-z)P^{(1,2)}_{\Theta}(z)\biggr\},  \\ \nonumber
P^{(1)}_{\Delta} &=& 2\ln\Delta + \frac{3}{2}\, , \quad
P^{(1)}_{\Theta}(z) = \frac{1+z^2}{1-z}\, ,\quad 
P^{(2)}_{\Delta} = \biggl(2\ln\Delta + \frac{3}{2}\biggr)^2 
- \frac{2\pi^2}{3}\, ,  \\ \nonumber
P^{(2)}_{\Theta}(z) &=& 2\biggl[ \frac{1+z^2}{1-z}\biggl(
2\ln(1-z)-\ln z + \frac{3}{2}\biggr) + \frac{1+z}{2}\ln z - 1 + z
\biggr], \\ \nonumber
{\cal D}(z) &=& \delta(1-z){\cal D}_{\Delta} 
+ \Theta(1-\Delta-z){\cal D}_{\Theta}(z), \\ \nonumber
{\cal D}_{\Delta} &=& 1 + {\cal D}_{\Delta}^{[\alpha]} 
+ {\cal D}_{\Delta}^{[\alpha^2]}  
+ \dots \ ,  \qquad
{\cal D}_{\Theta}(z) =  {\cal D}_{\Theta}^{[\alpha]}(z) 
+ {\cal D}_{\Theta}^{[\alpha^2]}  
+ \dots \ ,  \\ \nonumber
{\cal D}_{\Delta}^{[\alpha]} &=& \frac{\alpha}{2\pi}(L-1)
P^{(1)}_{\Delta},\quad
{\cal D}_{\Theta}^{[\alpha]}(z) = \frac{\alpha}{2\pi}(L-1)
P^{(1)}_{\Theta}(z),
\\ \nonumber
{\cal D}_{\Delta}^{[\alpha^2]} &=&
\biggl(\frac{\alpha}{2\pi}\biggr)^2\frac{(L-1)^2}{2!}
P^{(2)}_{\Delta},\quad
{\cal D}_{\Theta}^{[\alpha^2]}(z) =
\biggl(\frac{\alpha}{2\pi}\biggr)^2\frac{(L-1)^2}{2!}\;
P^{(2)}_{\Theta}(z), \qquad
L = \ln\frac{s}{m_e^2}\, .
\end{eqnarray}
Here $z$ means the energy fraction of an electron just before it emitted
a collinear photon. The leading lograthmic corrections due 
to electron--positron pair production can be easily added 
within the same formalism~\cite{JHEP97a,SF}.

\subsection{The kinematical regions}

Our idea is to decompose the formula according to different
types of the final state kinematics. Each part of the decomposition
is a particular contribution to the total Bhabha cross section, and
it can be measured, in principle, independently. So, as a result we
have a sum of positive quantities, which is important for Monte
Carlo simulations. The 13 contributions are given below. 

\begin{enumerate}

\item{}
{\bf The (quasi--)elastic kinematics.} 

Here we take into account the Born cross section with
virtual loop corrections and soft photons. The energy of soft
photons does not exceed $\Delta\varepsilon$. The parameter
$\Delta\ll 1$ is auxiliary, the final result (the total sum)
should not depend on its value. That provides an additional check
of our calculation. 

\begin{eqnarray}
\frac{\dd\sigma_1}{\dd\varphi\dd c} = 
\frac{\dd\widetilde{\sigma}(1,1)}{\dd\varphi\dd c} 
\biggl\{
1 + \frac{\alpha}{\pi}K_{SV} + 4{\cal D}_{\Delta}^{[\alpha]}
+ 6({\cal D}_{\Delta}^{[\alpha]})^2
+ 4{\cal D}_{\Delta}^{[\alpha^2]} 
- \frac{8\alpha}{\pi}\ln\Delta\,\ln(\mathrm{cot}(\frac{\theta}{2}))
\biggr\}.
\end{eqnarray} 

\item{$\vecc{k}\parallel\vecc{p}_1$:~}
{\bf one (two) photons along the initial electron momentum.}

In this case we observe emission of a hard collinear photon
inside a narrow cone
along the direction of motion of the initial electron. The
auxiliary parameter $m_e/\varepsilon\ll\theta_0\ll 1$ 
defines the cone: $\widehat{\vecc{k}\vecc{p}_1} < \theta_0$.
We suppose that the parameter is less than the angular resolution
of the detector. And therefore we are not going distinguish
the situations of one or two photon emission in this cone. We
just sum up the energies and momenta, if they are two.

\begin{eqnarray}
\frac{\dd\sigma_2}{\dd\varphi\dd c} =
\int\limits_{\bar{z}_1}^{1-\Delta}\dd z_1\; 
\frac{\dd\widetilde{\sigma}(z_1,1)}{\dd\varphi\dd c} 
\biggl\{ {\cal D}_{\Theta}^{[\alpha]}(z_1)
+ C_{\mathrm{ini}}(z_1)
+ 3{\cal D}_{\Theta}^{[\alpha]}(z_1){\cal D}_{\Delta}^{[\alpha]}
+ {\cal D}_{\Theta}^{[\alpha^2]}(z_1)
\biggr\}, 
\end{eqnarray} 
The lowest value of $z_1$ to be defined from the conditions
of particle registration. If $y_{\mathrm{th}}$ is the threshold
energy for electron registration, then
\begin{eqnarray}
\bar{z}_1 = \frac{y_{\mathrm{th}}(1+c)}{2-y_{\mathrm{th}}(1-c)}\, .
\end{eqnarray} 

The compensator for the initial state radiation is
\begin{eqnarray}
C_{\mathrm{ini}}(z) = \frac{\alpha}{2\pi}\biggl[
\frac{1+z^2}{1-z}\ln\frac{\theta_0^2}{4} + 1 - z \biggr]. 
\end{eqnarray} 

\item{$\vecc{k}\parallel\vecc{p}_2$:~}
{\bf one (two) photons along the initial positron momentum.}

This case is completely analogous to the previous one.
\begin{eqnarray}
\frac{\dd\sigma_3}{\dd\varphi\dd c} =
\int\limits_{\bar{z}_2(z_1=1)}^{1-\Delta}\dd z_2\; 
\frac{\dd\widetilde{\sigma}(1,z_2)}{\dd\varphi\dd c} 
\biggl\{ {\cal D}_{\Theta}^{[\alpha]}(z_2)
+ C_{\mathrm{ini}}(z_2) 
+ 3{\cal D}_{\Theta}^{[\alpha]}(z_2){\cal D}_{\Delta}^{[\alpha]}
+ {\cal D}_{\Theta}^{[\alpha^2]}(z_2)
\biggr\}.
\end{eqnarray} 
The lowest limit for $z_2$ is defined by
\begin{eqnarray} \label{z2min}
\bar{z}_2 = \frac{z_1y_{\mathrm{th}}(1-c)}{2z_1-y_{\mathrm{th}}(1+c)}\, ,
\end{eqnarray} 
where one has to imply $z_1=1$.

\item{$\vecc{k}\parallel\vecc{p}_1'$:~}
{\bf one (two) photons along the final electron momentum.}

\begin{eqnarray}
\frac{\dd\sigma_4}{\dd\varphi\dd c} &=&
\frac{\dd\widetilde{\sigma}(1,1)}{\dd\varphi\dd c} 
\int\limits_{y_{\mathrm{th}}}^{1-\Delta}\dd y_1\; 
\biggl\{ {\cal D}_{\Theta}^{[\alpha]}(y_1)
+ C_{\mathrm{fin}}(y_1) 
+ 3{\cal D}_{\Theta}^{[\alpha]}(y_1){\cal D}_{\Delta}^{[\alpha]}
+ {\cal D}_{\Theta}^{[\alpha^2]}(y_1)
\biggr\}.
\end{eqnarray} 

The compensator for the final state radiation is
\begin{eqnarray}
C_{\mathrm{fin}}(y) = \frac{\alpha}{2\pi}\biggl[
\frac{1+y^2}{1-y}\biggl( \ln\frac{\theta_0^2}{4} + 2\ln y \biggr)
+ 1 - y \biggr]. 
\end{eqnarray} 

\item{$\vecc{k}\parallel\vecc{p}_2'$:~}
{\bf one (two) photons along the final positron momentum.}

\begin{eqnarray}
\frac{\dd\sigma_5}{\dd\varphi\dd c} &=&
\frac{\dd\widetilde{\sigma}(1,1)}{\dd\varphi\dd c} 
\int\limits_{y_{\mathrm{th}}}^{1-\Delta}\dd y_2\; 
\biggl\{ {\cal D}_{\Theta}^{[\alpha]}(y_2)
+ C_{\mathrm{fin}}(y_2) 
+ 3{\cal D}_{\Theta}^{[\alpha]}(y_2){\cal D}_{\Delta}^{[\alpha]}
+ {\cal D}_{\Theta}^{[\alpha^2]}(y_2)
\biggr\}.
\end{eqnarray} 

\item{$\vecc{k}_a\parallel\vecc{p}_1$, $\vecc{k}_b\parallel\vecc{p}_2$:~}
{\bf one photon along the initial electron momentum
and one photon along the initial positron momentum.}

\begin{eqnarray}
\frac{\dd\sigma_6}{\dd\varphi\dd c} &=&
\int\limits_{\bar{z}_1}^{1-\Delta}\dd z_1\; 
\int\limits_{\bar{z}_2}^{1-\Delta}\dd z_2\; 
\frac{\dd\widetilde{\sigma}(z_1,z_2)}{\dd\varphi\dd c} 
{\cal D}_{\Theta}^{[\alpha]}(z_1) {\cal D}_{\Theta}^{[\alpha]}(z_2). 
\end{eqnarray} 

\item{$\vecc{k}_a\parallel\vecc{p}_1'$, $\vecc{k}_b\parallel\vecc{p}_2'$:~}
{\bf one photon along the final electron momentum
and one photon along the final positron momentum.}

\begin{eqnarray}
\frac{\dd\sigma_7}{\dd\varphi\dd c} &=&
\frac{\dd\widetilde{\sigma}(1,1)}{\dd\varphi\dd c} 
\int\limits_{y_{\mathrm{th}}}^{1-\Delta}\dd y_1\; 
\int\limits_{y_{\mathrm{th}}}^{1-\Delta}\dd y_2\; 
{\cal D}_{\Theta}^{[\alpha]}(y_1) {\cal D}_{\Theta}^{[\alpha]}(y_2). 
\end{eqnarray} 

\item{$\vecc{k}_a\parallel\vecc{p}_1$, $\vecc{k}_b\parallel\vecc{p}_1'$:~}
{\bf one photon along the initial electron momentum
and one photon along the final electron momentum.}

\begin{eqnarray}
\frac{\dd\sigma_8}{\dd\varphi\dd c} &=&
\int\limits_{\bar{z}_1}^{1-\Delta}\dd z_1\; 
\frac{\dd\widetilde{\sigma}(z_1,1)}{\dd\varphi\dd c} 
{\cal D}_{\Theta}^{[\alpha]}(z_1) 
\int\limits_{y_{\mathrm{th}}/Y_1}^{1-\Delta}\dd y_1\; 
{\cal D}_{\Theta}^{[\alpha]}(y_1), \\ \nonumber
Y_1 &=& \frac{2z_1}{z_1+1-c(z_1-1)}\, .
\end{eqnarray} 

\item{$\vecc{k}_a\parallel\vecc{p}_1$, $\vecc{k}_b\parallel\vecc{p}_2'$:~}
{\bf one photon along the initial electron momentum
and one photon along the final positron momentum.}

\begin{eqnarray}
\frac{\dd\sigma_9}{\dd\varphi\dd c} &=&
\int\limits_{\bar{z}_1}^{1-\Delta}\dd z_1\; 
\frac{\dd\widetilde{\sigma}(z_1,1)}{\dd\varphi\dd c} 
{\cal D}_{\Theta}^{[\alpha]}(z_1) 
\int\limits_{y_{\mathrm{th}}/Y_2}^{1-\Delta}\dd y_2\; 
{\cal D}_{\Theta}^{[\alpha]}(y_2), \\ \nonumber
Y_2 &=& \frac{z_1^2+1-c(z_1^2-1)}{z_1+1-c(z_1-1)}\, .
\end{eqnarray} 

\item{$\vecc{k}_a\parallel\vecc{p}_2$, $\vecc{k}_b\parallel\vecc{p}_1'$:~}
{\bf one photon along the initial positron momentum
and one photon along the final electron momentum.}

\begin{eqnarray}
\frac{\dd\sigma_{10}}{\dd\varphi\dd c} &=&
\int\limits_{\bar{z}_2}^{1-\Delta}\dd z_2\; 
\frac{\dd\widetilde{\sigma}(1,z_2)}{\dd\varphi\dd c} 
{\cal D}_{\Theta}^{[\alpha]}(z_2) 
\int\limits_{y_{\mathrm{th}}/Y_1}^{1-\Delta}\dd y_1\; 
{\cal D}_{\Theta}^{[\alpha]}(y_1), \\ \nonumber
Y_1 &=& \frac{2z_2}{z_2+1-c(1-z_2)}\, .
\end{eqnarray} 

\item{$\vecc{k}_a\parallel\vecc{p}_2$, $\vecc{k}_b\parallel\vecc{p}_2'$:~}
{\bf one photon along the initial positron momentum
and one photon along the final positron momentum.}

\begin{eqnarray}
\frac{\dd\sigma_{11}}{\dd\varphi\dd c} &=&
\int\limits_{\bar{z}_2}^{1-\Delta}\dd z_2\; 
\frac{\dd\widetilde{\sigma}(1,z_2)}{\dd\varphi\dd c} 
{\cal D}_{\Theta}^{[\alpha]}(z_2) 
\int\limits_{y_{\mathrm{th}}/Y_2}^{1-\Delta}\dd y_2\; 
{\cal D}_{\Theta}^{[\alpha]}(y_2), \\ \nonumber
Y_2 &=& \frac{z_2^2+1-c(1-z_2^2)}{z_2+1-c(1-z_2)}\, .
\end{eqnarray}

\item{$\vecc{p}_1'\parallel\vecc{p}_2'$:~}
{\bf both the final electron and positron go together 
back--to--back to a hard photon.}

This contribution works only if no any cut--off on acollinearity is
imposed. Its kinematics is just the one of the process of annihilation into
two photons (one of which is converted then into an electron--positron
pair with low invariant mass).
It gives a large logarithm, when the angle between the pair 
components is small.

\begin{eqnarray} \label{c12}
\dd\sigma_{12} &=& \dd\varphi\dd c
\int\limits_{y_{\mathrm{th}}}^{1-y_{\mathrm{th}}}\dd z\; 
\frac{\alpha^3}{2\pi s}\, \frac{1+c^2}{1-c^2}
\biggl( \ln\Delta_1 + L - 2\ln2 - \frac{5}{3} \biggr)
(z^2+(1-z)^2),
\end{eqnarray}
where $z$ denotes the energy fraction of the final electron,
the new auxiliary parameter $\Delta_1$ bounds the energy
of the hard photon for this contribution: 
$1>\omega > \varepsilon(1-\Delta_1)$. The opposite condition
is to be implied in the last ($WT$) contribution in order
to cancel out $\Delta_1$.

\item{}
{\bf One hard photon at large angle in respect to all other
particle momenta}

\begin{eqnarray} \label{c13}
\frac{\dd\sigma_{13}}{\dd\varphi\;\dd c} &=& 
\frac{\alpha^3}{2\pi^2 s}
\int\limits_{\Delta}^{1-\Delta_1}\dd x\;
\int\limits_{-1}^{1}\dd c_2\;
\int\limits_{0}^{2\pi}\dd\varphi_2\;
\frac{WT}{4} \Gamma' \Theta.
\end{eqnarray}
Letter $\Theta$ denotes the general restrictions on the
phase space: polar angles of the photon in respect to
any other momentum should be more than $\theta_0$; it
should also include the cut-off on acollinearity for the
final charged particles, if it is imposed.
\begin{eqnarray}
W &=& \frac{s}{\chi_+\chi_-} + \frac{s_1}{\chi'_+\chi'_-} 
- \frac{t}{\chi_-\chi'_-} - \frac{t_1}{\chi_+\chi'_+} 
+ \frac{u}{\chi_-\chi'_+} + \frac{u_1}{\chi_+\chi'_-}\, , \\ \nonumber
T &=& \frac{ss_1(s^2+s_1^2)+tt_1(t^2+t_1^2)+uu_1(u^2+u_1^2)}
{ss_1tt_1}\, ,\\ \nonumber
s &=& 4, \quad t = - 2Y_1(1-c), \quad u = - 2Y_2(1+c_3), 
\\ \nonumber
s_1 &=& 2Y_1(2-x+xc_1)=4(1-x), \quad t_1 = - 2Y_2(1-c_3), \quad 
u_1 = - 2Y_1(1+c), 
\\ \nonumber
\chi_- &=& x(1-c_2), \quad \chi_+ = x(1+c_2), \quad 
\chi'_- = xY_1(1-c_1), \quad \chi'_+ = xY_2(1-c_5),  
\\ \nonumber
Y_1 &=& \frac{2(1-x)}{2-x(1-c_1)}\, , \qquad
Y_2 = 2 - Y_1 - x,  \\ \nonumber 
\Gamma' &=& \frac{\dd\Gamma}{\varepsilon^2\dd\varphi\;
\dd c\;\dd\varphi_2\;\dd c_2\;\dd x} = \frac{xY_1}{2-x+xc_1}\, .
\end{eqnarray}

The current version of the program contains the option to take into account
the vacuum polarization affect in this contribution~\cite{JHEP97a}.

\end{enumerate}

The complete final state kinematics can be defined in each
contribution according to the general formulae given 
in Ref.~\cite{JHEP97a}.

\section{Event generator structure}

An original algorithm for event generation was applied in the code.
The steps are as follows.

\begin{itemize}

\item[]{\bf Step 1.}
At first we perform the numerical integration of the 13 contributions 
over the phase space, where only the most general cuts are applied:
\begin{eqnarray}
\sigma_i = \int\frac{\dd\sigma_i}{\dd\Gamma_i}\;\dd\Gamma_i, \qquad 
i = 1,\; \dots\; , 13.
\end{eqnarray}
In this way we obtain the relative weights of the 13 contributions
and at the same moment the absolute value of the total cross section:
\begin{eqnarray}
U_i = \frac{\sigma_i}{\sigma_{\mathrm{tot}}}, \qquad
\sigma_{\mathrm{tot}} = \sum_{i=1}^{13} \sigma_i.
\end{eqnarray}

\item[]{\bf Step 2.}
The total ordered number of events to be generated $N_{\mathrm{tot}}$ 
is shared between
the 13 contributions according their relative contributions to the
total cross section. That is done as follows. The mail subroutine of
the generator is called $N_{\mathrm{tot}}$ times. Each time we choose one of the 13
kinematical regions according to the value of a random number $r$, by
comparing it with the relative weights $w_i$:
\begin{eqnarray}
{\mathrm{if}}\quad \sum_{k=1}^{j}U_k < r \leq  \sum_{k=1}^{j+1}U_k, \quad
{\mathrm{then}} \quad i = j; \\ \nonumber
{\mathrm{if}}\quad r \leq U_1,\quad {\mathrm{then}} \quad i = 1.
\end{eqnarray}

\item[]{\bf Step 3.}
Now a set of kinematical variables $v_n$ for an event of the chosen 
contribution
is generated. For each particular differential distribution 
$\dd\sigma_i/\dd\Gamma_i$ we use a specific change of variables in
order to make the distribution more flat. The weight of the event is
defined by the formula: 
\begin{eqnarray}
w_n = \frac{\dd\sigma_i(v_n)}{\dd\Gamma_i}\;
\frac{N_{\mathrm{tot}}}{\sigma_{\mathrm{tot}}}\, .
\end{eqnarray}
Calculating the value of $\dd\sigma_i$ we apply the same cuts and 
conditions as while the numerical integration.

\item[]{\bf Step 4.} Because the differential distributions are
rather complicated, we were not able to find a substitution to
make them completely flat. So, the weights, obtained in the third
step, can be different from unit. Here we do the following trick:
generate a random number $r$ and define the number of corresponding
unweighted events as the integer part of $w_n + 1 - r$.
If the obtained number $m_n$ is more than 1, we use the rotation
symmetry and distribute the events uniformly in the polar angle
of the scattered electron. 

\item[]{\bf Step 5.} At this step we can analyse the events generated
in the previous step and apply additional cuts, if required. We can
also record the events for future processing. 

\item[]{\bf Step 6.} After we executed the steps 2--4 $N_{\mathrm{tot}}$ times,
we can compare the results of numerical and Monte Carlo procedures.
That provides a control of the technical precision of the code.
Namely, we compare the total number of unweighted events with the
ordered one: 
\begin{eqnarray}
\sum_{n=1}^{N_{\mathrm{tot}}}m_n \approx N_{\mathrm{tot}}.
\end{eqnarray}
At the same moment this means that the value of the cross section
for the generated events is close to the one obtained by the numerical
integration. Such a comparison is also done for each of
13 contributions separately. The technical precision can be improved
by increasing of the total number of events and also by tuning parameters
of the program for a concrete task.

\end{itemize}

\section{Flags and parameters}

The reading of flags and parameters is performed by means of the
standard {\tt FFREAD} subroutine, which is called from the 
{\tt PACKLIB}~\cite{packlib}.

The code contains several flags which can switch between different
options in the physical base and in the generation procedure. Below
we describe the most important flags.

\begin{tabbing}
\= {\tt ICOR}=0:~~~\= calculations only at the Born level  \kill
\>  \underline{{\tt ICOR}} \>
                       \\
\> {\tt ICOR}=0:\> calculations only at the Born level    
                       \\
\> {\tt ICOR}=1:\> Born + LLA corrections
                       \\
\> {\tt ICOR}=2:\> Born + LLA corrections + $K$-factor
                       \\
\> {\tt ICOR}=3:\> Born + LLA corrections + $K$-factor + large-angle photon
                       \\[.2cm]
\>  \underline{{\tt IORD}} \>
                       \\
\> {\tt IORD}=0:\> calculations only at the Born level    
                       \\
\> {\tt IORD}=1:\> ${\cal O}(\alpha)$ RC are taken into account
                       \\
\> {\tt IORD}=2:\> higher order LLA corrections are included
                       \\[.2cm]
\>  \underline{{\tt IVPOL}} \>
                       \\
\> {\tt IVPOL}=0:\> $\alpha_{QED}$ is not running
                       \\
\> {\tt IVPOL}=1:\> vacuum polarization by leptons and hadrons is accounted
                       \\[.2cm]
\end{tabbing}
The $\Phi$-meson contribution to the photon virtual propagator is realized
in the program as a part of the vacuum polarization function. This option
can be switched on/off by the flag {\tt IPHI}=1/0.

The main parameters to be set by user are:
\begin{tabbing}
\= {\tt EB}:~~~~\= the beam energy in GeV;
                       \\
\> {\tt TETN}:\> the minimal electron scattering angle in radian;
                       \\
\> {\tt TETX}:\> the maximal electron scattering angle in radian;
                       \\
\> {\tt TEPN}:\> the minimal positron scattering angle in radian;
                       \\
\> {\tt TEPX}:\> the maximal positron scattering angle in radian;
                       \\
\> {\tt NEVE}:\> the number of events to be generated;
                       \\
\> {\tt TACO}:\> the minimal allowed angle between the outgoing electron
                 and positron in radian;
                       \\[.2cm]
\end{tabbing}

In order to take into account vacuum polarization in the 13th contribution
one has to set {\tt IEWT}=1 and {\tt IVWT}=1. This corrections is a part of
the second order next--to--leading contributions 
$(\sim {\cal O}(\alpha^2L))$, but it might be really important 
in the region close to resonance peaks.

\section{Conclusions}

The presented formulae are valid for the electron--positron
colliders of moderately high energies below 3~GeV. In order to
expand them for higher energies we take into account $Z$-boson
exchange and the relevant electroweak radiative corrections~\cite{EW},
but the corresponding option of the code will be described elsewhere.

In Table~1 we give the results for a rather simple configuration:
beam energy 0.5~GeV, scattering angles for the both positron
and electron lie between $15^{\circ}$ and $165^{\circ}$,
minimal angle between the final particles $30^{\circ}$,
threshold for electron registration 50~MeV. For the Table we
generated $10^8$ unweighted events.
\begin{table}[ht]
\caption{Results of numerical calculations}
\begin{tabular}[]{|l|c|c|c|c|c|}
\hline
 & $\sigma_{\mathrm{Born}}$ & 
$\sigma^{(1)}_{\mathrm{LLA}}$ & $\sigma^{(1)}_{\mathrm{LLA+K}}$ &
$\sigma^{(1)}$ & $\sigma^{(2)}$  \\
\hline
 & \multicolumn{5}{c|}{without vacuum polarization} \\ \hline
Numer. int., mbarn & 13.501 & 13.360 & 13.023 & 13.325 & 13.303     
\\ \hline
Monte Carlo, mbarn & 13.501 & 13.359 & 13.024 & 13.324 & 13.298   
\\ \hline
$\delta_{\mathrm{RC}}$, \% 
                   & ----- & $-1.05$ &$-3.55$ &$-1.31$ & $-1.47$    
\\ \hline
 & \multicolumn{5}{c|}{with vacuum polarization} \\ \hline
Numer. int., mbarn & 13.749 & 13.604 & 13.261 & 13.631 & 13.638 
\\ \hline
Monte Carlo, mbarn & 13.749 & 13.603 & 13.259 & 13.627 & 13.632 
\\ \hline
$\delta_{\mathrm{RC}}$, \% 
                   & 1.83   &  0.76  &$-1.78$ &  0.96  & $1.01$
\\ \hline
\end{tabular}
\end{table} 
The last line of the Table represents the relative difference
of the cross section in the corresponding approximation in respect to
the Born one:
\begin{eqnarray}
\delta_{\mathrm{RC}} = \frac{\sigma_i-\sigma_{\mathrm{Born}}}
{\sigma_{\mathrm{Born}}}\; 100\%.
\end{eqnarray}
Further, in the Table we denote: $\sigma^{\mathrm{Born}}$ is the pure
Born level QED cross section; 
$\sigma^{(1)}_{\mathrm{LLA}}$ is the
cross section with the ${\cal O}(\alpha)$ LLA corrections;
in $\sigma^{(1)}_{\mathrm{LLA+K}}$ the ${\cal O}(\alpha)$ $K$-factor
 due to soft and virtual photons is included; quantity 
$\sigma^{(1)}$ includes the complete ${\cal O}(\alpha)$
set of corrections; 
the complete set of the first order corrections
plus the LLA second order effects $\sigma^{(2)}$ 
are presented in the last column.

The resulting precision of the code for the description of large--angle
Bhabha scattering in the typical conditions of electron--positron
colliders of energy about a few GeVs is estimated 
to be 0.2\%~\cite{JHEP97a}. The uncertainty
can be decreased. In particular, an extended program for calculations
of second order next--to--leading RC to large--angle Bhabha scattering
is in progress~\cite{LABS}. From the other hand, the uncertainty
should be re--estimated  taking into account the concrete
experimental conditions and comparisons with other codes.

Comparisons with other available Monte Carlo event generators
for Bhabha scattering are in progress. 
The next step of the code development will be the explicit 
generation of radiated pairs. The third order leading logarithmic
corrections will be also taken into account. 
The current version of the program does not include the strange
effect of a diviation from the leading logarithmic approximation
in the ${\cal O}(\alpha^2L^2)$ order,
which was recently found in Ref.~\cite{AKS}.

The presented version of the code is dealing with the
large--angle Bhabha scattering. That does not mean that the code
can not be used for small--angles, but that it does not include
the complete set of the second--order next--to--leading corrections, 
which are known
only for the small--angle limiting case~\cite{SABS}. An extended 
version of the code, which will provide also the small--angle Bhabha
event generation, to be described elsewhere.

\subsection*{Acknowledgments}

Valuable discussions with G.~Fedotovich, 
E.~Kuraev, L.~Trentadue, and B.Shaikhatdenov
helped me very much. I am grateful also for support to the INTAS 
foundation, grant 93--1867 ext.


\begin{thebibliography}{99}

\bibitem{Bhabha}
H.J. Bhabha,
Proc. Roy. Soc. (London) {\bf A 154} (1935) 195.

\bibitem{LEP1}
H. Anlauf {\it et al.}, S. Jadach, O. Nicrosini (convenors),
CERN Yellow Report 96-01, vol. 2, 1996, p.229; \\
A.B. Arbuzov {\it et al.},
Phys. Lett. {\bf B383} (1996) 238.

%
\bibitem{l3}
G. Montagna {\it et al.},
Nucl. Phys. {\bf B 401} (1993) 3.

\bibitem{l4}
W. Placzek {\it et al.},
preprint CERN-TH/99-07, Geneva, 1997; hep-ph/9903381.

\bibitem{l5}
M.~Caffo, R.~Gatto, E.~Remiddi, 
Nucl. Phys. {\bf B 252} (1985) 378.

\bibitem{fed}
A.B.~Arbuzov, G.V.~Fedotovich, E.A. Kuraev, 
in preparation.

\bibitem{JHEP97a}
A.B.~Arbuzov {\it et al.}, JHEP {\bf 10} (1997) 001.

\bibitem{SF}
E.A. Kuraev, V.S. Fadin, 
Sov. J. Nucl. Phys. {\bf 41} (1985) 466.

\bibitem{EW}
W. Beenakker, F.A. Berends, S.C. van der Marck,
Nucl. Phys. {\bf B349} (1991) 323; \\
M. B\"ohm, A. Denner, and W. Hollik,
Nucl. Phys. {\bf B 304} (1988) 687; \\
F.A. Berends, R. Kleiss, W. Hollik,
Nucl. Phys. {\bf B 304} (1988) 712.

\bibitem{packlib}
R. Brun {\it et al.},
{\tt FFREAD} User Guide and Reference Manual,
CERN DD/US/71, CERN Program Library I 302, February 1987.

\bibitem{LABS}
A.B.~Arbuzov {\it et al.}, Nucl. Phys. {\bf B 474} (1996) 271;
{\it ibid} {\bf B 483} (1997) 83;
Phys. Atom. Nucl. {\bf 60} (1997) 591;
Mod. Phys. Lett. {\bf A 13} (1998) 2305;
hep--ph/9805308.

\bibitem{AKS}
V.~Antonelli, E.A.~Kuraev, B.G.~Shaikhatdenov,
hep--ph/9905331. 

\bibitem{SABS}
A.B.~Arbuzov {\it et al.}, 
Nucl. Phys. {\bf B 485} (1997) 457.


\end{thebibliography}
\end{document}